\documentclass[12pt,a4paper]{article}
\usepackage[]{amsmath,amssymb}
\usepackage{graphicx}
\usepackage{geometry}
\usepackage{amssymb,epsfig,subfigure}
\usepackage{graphics,epsfig}
\usepackage{graphicx, color}
\usepackage{geometry}
\usepackage{diagbox}
\usepackage{float}
\usepackage{xcolor}
\usepackage{bbold}
\usepackage[latin1]{inputenc}
\usepackage[font=footnotesize]{caption}
\usepackage[T1]{fontenc}
\usepackage{amsmath}
\usepackage{amsfonts}
\usepackage{amssymb}
\usepackage{latexsym}
\usepackage[english]{babel}
\usepackage{authblk}
\usepackage{slashed}
\newcommand{\be}{\begin{equation}}
\newcommand{\ee}{\end{equation}}
\newcommand{\beq}{\begin{equation}}
\newcommand{\eeq}{\end{equation}}
\newcommand{\bea}{\begin{eqnarray}}
\newcommand{\eea}{\end{eqnarray}}

\def\be{\begin{equation}}
\def\ee{\end{equation}}
\def\ba{\begin{eqnarray}}
\def\ea{\end{eqnarray}}

\definecolor{princetonorange}{rgb}{1.0, 0.56, 0.0}
\definecolor{WildStrawberry}{rgb}{1.0, 0.26, 0.64}
\definecolor{rossocorsa}{rgb}{0.83, 0.0, 0.0}
\definecolor{navyblue}{rgb}{0.0, 0.0, 0.5}
\usepackage{hyperref}
\hypersetup{
    colorlinks,
    citecolor=rossocorsa,
    filecolor=navyblue,
    linkcolor=navyblue,
    urlcolor=navyblue
}

\begin{document}
\title{Modular Hamiltonian in the semi infinite line, Part II: dimensional reduction of Dirac fermions in spherically symmetric regions}
%\title{Refinements, observations on entanglement entropy for gauge fields}

\author{Marina Huerta\footnote{e-mail: marina.huerta@cab.cnea.gov.ar} }
\author{Guido van der Velde\footnote{e-mail: guido.vandervelde@ib.edu.ar }}
\affil{Centro At\'omico Bariloche, 8400-S.C. de Bariloche, R\'{\i}o Negro, Argentina}

\date{}

\maketitle
\begin{abstract}
In this article, we extend our study on a new class of modular Hamiltonians on an interval attached to the origin on the semi-infinite line, introduced in a recent work dedicated to scalar fields. Here, we shift our attention to fermions and similarly to the scalar case, we investigate the modular Hamiltonians of theories which are obtained through dimensional reduction, this time, of a free massless Dirac field in $d$ dimensions.
By following the same methodology, we perform dimensional reduction on both the physical and modular Hamiltonians. This process enables us to establish a correspondence: we identify the modular Hamiltonian in an interval connected to the origin to the one obtained from the reduction of the modular Hamiltonian pertaining to the conformal parent theory on a sphere. Intriguingly, although the resulting one-dimensional theories lack conformal symmetry due to the presence of a term proportional to $1/r$, the corresponding modular Hamiltonians are local functions in the energy density. This phenomenon mirrors the well known behaviour observed in the conformal case and indicates the existence of a residual symmetry, characterized by a subset of the original conformal group.

Furthermore, through an analysis of the spectrum of the modular Hamiltonians, we derive an analytic expression for the associated entanglement entropy. Our findings also enable us to successfully recover the conformal anomaly coefficient from the universal piece of the entropy in even dimensions, as well as the universal constant $F$ term in $d=3$, by extending the radial regularization scheme originally introduced by Srednicki to perform the sum over angular modes.
\end{abstract}

\newpage

\tableofcontents

\section{Introduction} \label{intro}
In recent years, there has been a significant amount of research focused on the study of modular Hamiltonians, attracting interest across various research domains. This heightened attention stems mainly from two factors: their direct connection to entanglement measures and their fundamental role within the algebraic approach to quantum field theory (QFT) \cite{Haag}.

The modular Hamiltonian can be introduced through its explicit relation with reduced states associated with a local algebra of operators within a region. In general, these reduced states can be conveniently expressed (with the presence of a cutoff) as a density matrix given by
\begin{equation}
\rho = \frac{e^{-K}}{\textrm{tr}e^{-K}},
\label{thermal}
\end{equation}
where $K$ represents the modular Hamiltonian operator. Remarkably, equation (\ref{thermal}) establishes a connection between the entanglement entropy and the thermodynamic entropy of a system in equilibrium at temperature $1$ with respect to the modular Hamiltonian $K$.

Our current understanding of the explicit form of modular Hamiltonians is mostly limited to a few examples where the modular flow is local and primarily determined by spacetime symmetries. One such example is the Rindler wedge $x^1 > \vert t \vert$ in Minkowski space, applicable to any QFT, which can be expressed as
\begin{equation}
\rho = k\, e^{-2\pi \int_{x^1>0}d^{d-1}x\,x^1 T_{00}(x)},
\label{rindler}
\end{equation}
where $K$ is identified as $2\pi$ times the generator of boosts restricted to act only on the right Rindler wedge
\begin{equation}
K = 2\pi \int_{x^1>0}d^{d-1}x~x^1 T_{00}(x).
\end{equation}

Remarkably, equation (\ref{rindler}) reveals that the vacuum state in a half-space can be interpreted as a thermal state with an inverse temperature of $2\pi$ with respect to the boost operator. Another well-known example where symmetries play a crucial role in determining the exact modular Hamiltonian is that of conformal field theories (CFT) defined on spheres in any dimension
\begin{equation}
K = 2\pi \int_{|\vec{x}|<R} d^{d-1}x \frac{R^2-r^2}{2R} T_{00}(\vec{x}),
\label{modesf}
\end{equation}
which corresponds to the transformed Rindler modular Hamiltonian resulting from a conformal transformation mapping the Rindler wedge to causal spherical regions. This transformation leaves the vacuum of a CFT invariant.

In equation (\ref{modesf}), $K$ remains local and proportional to $T_{00}$, with a weight function 
\begin{equation}
\beta(r) = \frac{R^2-r^2}{2R}
\end{equation} 
serving as the proportionality factor. Apart from these examples, only a few other modular Hamiltonians, both local and non-local, are known. However, from a quantum information perspective, one does not generally expect locality to hold. In general, $K$ will be a non-local and non-linear combination of field operators at different positions within the region. An explicit example of a non-local modular Hamiltonian is derived for the vacuum state of a free massless fermion in $d = 2$ for several disjoint intervals \cite{Casini:2009vk,Longo:2009mn,Wong:2018svs}. In this case, $K$ consists of a local term proportional to the energy density and an additional non-local part given by a quadratic expression in the fermion field that connects points located in different intervals.

More recently, a new class of local modular Hamiltonians for non-conformal theories defined in the radial half-line, and corresponding to an interval attached to the origin $(0,R)$, was introduced in \cite{Huerta:2022tpq}. This new result involved dimensionally reducing the free massless scalar theory, originally in arbitrary $d$ spacetime dimensions. Building upon this work, we extend this class to the case where the parent theory is a massless Dirac fermion. The strategy employed follows the same approach as in \cite{Huerta:2022tpq}, where the modular Hamiltonian of the reduced system is calculated by leveraging the known modular Hamiltonian (\ref{modesf}) of the parent CFT, massless scalars in \cite{Huerta:2022tpq} and Dirac fermions in our case, on spheres in any dimension.

\begin{figure}[t]
\begin{center}  
\includegraphics[width=0.50\textwidth]{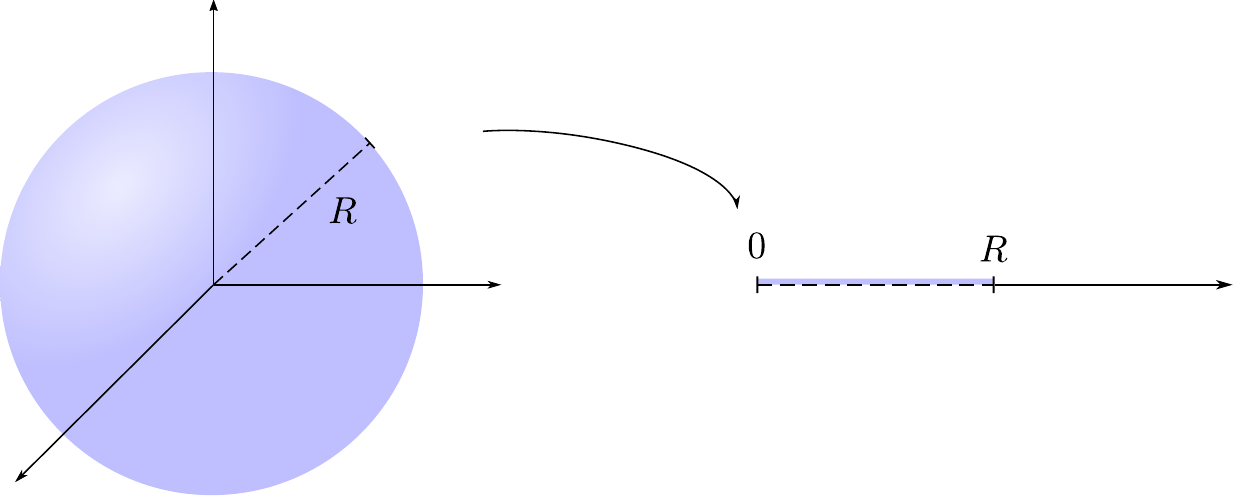}
\captionsetup{width=0.9\textwidth}
\caption{Sketch of the dimensional reduction: algebras assigned to spherical regions with radius $R$ correspond to operators supported in the segment $(0,R)$ attached to the origin.}
\label{sphere_interval}
\end{center}  
\end{figure}

The free massless Dirac fermion in $d$ spacetime dimensions can be dimensionally reduced to a sum of one-dimensional theories, each associated with an angular mode. In spherical coordinates, by introducing a field decomposition based on a complete orthonormal set of spherical spinor eigenfunctions of the Dirac operator, we integrate over the angular coordinates in the Hamiltonian density. Since the reduction is achieved by integrating over the angular coordinates, these systems naturally reside on the semi-infinite line. As mentioned in \cite{Huerta:2022tpq}, this angular integration is particularly convenient from an algebraic perspective when studying algebras assigned to spherical regions, such as when calculating entanglement entropy. In these coordinates, the local algebra associated with the region can be expressed easily in terms of fields $\Psi(r,\Omega)$ with desirable localization properties. For instance, points on the semi-infinite line correspond to shells in the original space, while intervals connected to the origin correspond to $d$-spheres (see Figure \ref{sphere_interval}).

In the radial coordinate, the canonical Hamiltonian for the massless free Dirac field decomposes as a sum over angular modes $H_{\ell m s}$:
\begin{equation}
H = \sum_{\ell m s} H_{\ell m s},
\end{equation}
where $\{\ell m s\}$ represents the angular mode label. Similarly, the modular Hamiltonian (\ref{modesf}) also decomposes into individual modes:
\begin{equation}
K = \sum_{\ell m s} K_{\ell m s}.
\end{equation}
Considering that the vacuum state for a system composed of independent subsystems is a product of density matrices, i.e., $\rho = \otimes \rho_{\ell m s}$, we can readily identify the modular Hamiltonian mode $K_{\ell m s}$ with the modular Hamiltonian of the one-dimensional reduced system $H_{\ell m s}$. However, $H_{\ell m s}$ does not correspond to a conformal relativistic theory due to an additional quadratic term proportional to $1/r$, where the proportionality constant depends on the dimension of the original problem and the angular mode $\ell$. Surprisingly, we find that $K_{\ell m s}$ remains local and proportional to the energy density $T_{00}$, similar to the scalar case.

The article is organized as follows: section \ref{sec: spherical_coordinates} explicitly carries out the dimensional reduction. By expressing the Dirac field in a basis of $(d-2)$-dimensional spinors, which are eigenfunctions of the Dirac operator in the angular directions, we integrate out the angular coordinates and obtain an action for the reduced system $S=\sum_{\ell m s}S_{\ell m s}$ of the form:
\begin{equation}\label{reducedfermion}
\begin{split}
S_{\ell m s}=i\int dt dr &\left\{\overline{\psi}_{+ \ell m}^{(s)}\left(\sigma^3\partial_t+i\sigma^2\partial_r -\frac{\lambda}{r}\sigma^1\right)\psi_{+ \ell m}^{(s)}\right.\\
&\left.+ \overline{\psi}_{-\ell m}^{(s)}\left(\sigma^3\partial_t+i\sigma^2\partial_r +\frac{\lambda}{r}\sigma^1\right)\psi_{- \ell m}^{(s)}\right\}\end{split}
\end{equation}
%\begin{equation}
%H_{nls} = \frac{1}{2} \int dr \left(XXXXXXX\right),
%\label{h1d}
%\end{equation}
with
\begin{equation}
\left\{\psi_{(\pm)\ell m}^{(s)}(r),\psi_{(\pm) \ell' m'}^{(s')\dagger}(r')\right\}=i\delta(r-r')\delta_{\ell'\ell}\delta_{m'm}\delta_{s's}.
\end{equation}
and
\begin{equation}
\lambda\equiv \ell+\frac{d-2}{2}\,.
\end{equation}
In section \ref{sec: modHam}, the same procedure is employed to find the modular Hamiltonian:
\begin{equation}
K_{\ell m s} = 2\pi \int_{0}^{R} dr \left(\frac{R^2-r^2}{2R}\right) T_{00}^{\ell m s}(r).
\label{Kintro}
\end{equation}
In a way, the reduced theory, which is manifestly invariant under dilatations but non-conformal, keeps trace of the conformal symmetry of the parent $d$-dimensional theory \cite{Jackiw:2011vz,Chamon:2011xk}, as it shares the same local modular Hamiltonian as the vacuum of a CFT in a sphere. In fact, we demonstrate that the modular transformation is an inherited conformal map that leaves the reduced theories invariant. The modular Hamiltonian (\ref{Kintro}), when expressed as a Noether charge restricted to the segment $(0,R)$, can be accordingly interpreted as the local operator implementing the modular flow.

In section \ref{sec: spectrum}, we solve the spectrum of the modular Hamiltonian (\ref{Kintro}), which is used in section \ref{sec: entropy} to compute the entanglement entropy in a segment connected to the origin. We derive the analytical expression:
\begin{equation}
S(\ell,d)=S_{+}(\ell,d)+S_{-}(\ell,d)\, ,
\end{equation}
with
\begin{equation}
S_{\pm}(\ell,d)=\frac{1}{6}\log{\frac{R}{\epsilon}}- i \pi \int_0^{\infty} ds \frac{s}{\cosh^2{\pi s}}\log{\left[\frac{2^{-2 i s}\Gamma\left(1+2 i s\right)\Gamma\left(-i s\right)\Gamma\left(\ell+\frac{d-1}{2}-i s\right)}{\Gamma\left(1-2 i s\right)\Gamma\left(i s\right)\Gamma\left(\ell+\frac{d-1}{2}+i s\right)}\right]}
\, ,
\end{equation}
which exhibits a logarithmic divergence with the expected coefficient $1/6$ (in $d=2$, there is only one contribution $S_{+}$).  It also contains a constant term that depends on both the mode $\ell$ and the spacetime dimension $d$ of the original theory. Although the above integral cannot generally be solved analytically, we make useful approximations to extract relevant information out of it. Furthermore, by summing over $\ell$, we recover the conformal anomaly in the logarithmic coefficient for the free Dirac field in even dimensions, as well as the constant universal $F$ term in $d=3$. To perform the sum over angular modes $\ell$, we introduce a novel regularization scheme using a damping exponential $\exp[-\left(\ell+1/2\right)\epsilon/R]$, where the cutoff $\epsilon$ is the same as that used to regularize the radial coordinate $r$. This regularization method generalizes the radial regularization scheme introduced by Srednicki for scalars fields in \cite{Srednicki:1993im} (also see \cite{Riera:2006vj}), which explicitly states that a radial lattice is insufficient for regularization when $d-1 \geqslant 4$ and the sum over partial waves fails to converge. In the case of fermions, this is so already for $d=4$ \cite{Benedetti:2023jye}. Finally, we conclude with some remarks.

\section{Spherical coordinates: dimensional reduction}\label{sec: spherical_coordinates}

A necessary ingredient for writing a free fermion action in spherical coordinates is choosing a tetrad basis and a spin connection, which enforce local Lorentz invariance. Furthermore, there is an ambiguity in the representation of the gamma matrices that satisfy the Clifford algebra. In this article we follow the conventions of \cite{Lopez-Ortega:2009flo} which, together with successive Weyl transformations, allow to decompose the Dirac operator into a direct product, one block representing the Dirac operator of a two dimensional spacetime in the radial direction, and the other corresponding to the transverse $d-2$ dimensional unit sphere. We briefly review the most important steps in the derivation of the dimensional reduction, explained in full detail in \cite{Lopez-Ortega:2009flo}.

The covariant form of the theory reads
\begin{equation}\label{action1}
S=i\int d^d x \sqrt{g} ~ \overline{\psi} \slashed{\nabla}\psi\, ,
\end{equation}
where the covariant derivative is defined in terms of the veilbein and spin connection
\begin{equation}
\slashed{\nabla}=\gamma^a e_a^{\mu}\left(\partial_{\mu}+\frac{1}{4}\omega_{\mu b c} \gamma^b\gamma^c \right)\, ,
\end{equation}
Latin and Greek letters representing flat space and spherical coordinates indices, respectively. We use the representation
\begin{equation}
\begin{split}
\gamma^1&=\sigma^3 \otimes \mathbb{1}_2\otimes \mathbb{1}_2\otimes \mathbb{1}_2\cdots=\sigma^3\otimes \mathbb{1}_{ 2^{\left[(d-2)/2\right]}}\, ,\\
\gamma^2&=i\sigma^2 \otimes \mathbb{1}_2\otimes \mathbb{1}_2\otimes \mathbb{1}_2\cdots=i\sigma^2\otimes \mathbb{1}_{ 2^{\left[(d-2)/2\right]}}\, ,\\
\gamma^3&=i\sigma^1\otimes \sigma^3 \otimes \mathbb{1}_2\otimes \mathbb{1}_2\cdots=i\sigma^1\otimes \hat{\gamma}^1\, ,\\
\gamma^4&=i\sigma^1\otimes \sigma^2 \otimes \mathbb{1}_2\otimes \mathbb{1}_2\cdots=i\sigma^1\otimes \hat{\gamma}^2\, ,\\
\gamma^5&=i\sigma^1\otimes \sigma^1 \otimes \sigma^3\otimes \mathbb{1}_2\cdots=i\sigma^1\otimes \hat{\gamma}^3\, ,\\
\vdots &\\
\gamma^d&=\cdots=i\sigma^1 \otimes \hat{\gamma}^{d-2}\, ,
\end{split}
\end{equation}
with $\hat{\gamma}^{d-2}$ a representation of the gamma matrices in a $d-2$ dimensional space with signature $(+,\cdots,+)$. 

Crucially, (\ref{action1}) is weyl invariant, so we can express it in a rescaled frame
\begin{equation}
\begin{split}
d\tilde{s}^2&=\frac{1}{r^2}d s^2=\frac{d t^2}{r^2}-\frac{d r^2}{r^2}-d\Sigma_{d-2}^2\, ,\\
\tilde{\psi}&=r^{(d-1)/2}\psi\, ,\\
\tilde{\slashed{\nabla}}\tilde{\psi}&=r^{(d+1)/2}\slashed{\nabla}\psi\, .
\end{split}
\end{equation}
The advantage of doing such a transformation is that, since the veilbein writes
\begin{equation}
\begin{split}
\tilde{e}^1 &=\frac{d t}{r}\, ,\\
\tilde{e}^2 &=\frac{d r}{r}\, ,\\
\tilde{e}^i &= h^i_j(\phi_k)d\phi^j\, ,
\end{split}
\end{equation}
with $\phi_k$ the angular coordinates, the matrix defining the one-form spin connection takes a block diagonal form, in which $\tilde{\omega}_{1 i}=\tilde{\omega}_{2 j}=0$. Therefore,
\begin{equation}\label{action2}
\begin{split}
S&=i\int d^d x \sqrt{\tilde{g}}~ \overline{\tilde{\psi}}\tilde{\slashed{\nabla}}\tilde{\psi}\\
&=i\int d^d x \sqrt{\tilde{g}}~ \overline{\tilde{\psi}}\left( \tilde{\slashed{\nabla}}_{d=2}\otimes \mathbb{1}_{ 2^{\left[(d-2)/2\right]}}+i\sigma^1\otimes \slashed{\nabla}_{\Sigma_{d-2}}\right)\tilde{\psi}\, .
\end{split}
\end{equation}
$\tilde{\slashed{\nabla}}_{d=2}$ is the Dirac operator corresponding to a two dimensional spacetime with metric 
\begin{equation}
d \tilde{s}_{d=2}^2=\frac{d t^2}{r^2}-\frac{d r^2}{r^2}\, ,
\end{equation}
whereas $\slashed{\nabla}_{\Sigma_{d-2}}=\hat{\gamma}^i\nabla_i$ is the Dirac operator  
on the $d-2$ dimensional unit sphere. Its eigenfunctions are well known \cite{Camporesi:1995fb}
\begin{equation}
\hat{\gamma}^i\nabla_i \chi_{(\pm)\ell m}^{(s)}(\Omega)=\pm i \left(\ell+\frac{d-2}{2}\right)\chi_{(\pm)\ell m}^{(s)}(\Omega)\, , \quad \ell \in \mathbb{N}_0\, .
\end{equation}
satisfying orthonormality relations 
\begin{equation}
\int_{S^{d-2}}d\Omega_{d-2} \chi_{(\pm)\ell m}^{(s)\dagger}(\Omega)\chi_{(\pm)\ell' m'}^{(s')}(\Omega)=\delta_{\ell'\ell}\delta_{m'm}\delta_{s's}.
\end{equation}
Here $m$ is a collective index for integer values, $\vec{m}=\{m_1,..., m_{d-3}\}$, satisfying $\ell\geq m_1\geq\ldots\geq m_{d-3}$. Meanwhile, for each set of angular momentum modes there are $2^{\left[ \frac{d-2}{2}\right]}$ distinct eigenfunctions, labelled by $s$. For this reason the degeneracy related to each eigenvalue is 
\begin{equation}
\lambda(\ell, d)=\frac{2^{\left[(d-2)/2\right]}(d+\ell-3)!}{\ell! (d-3)!}\, .
\end{equation}
Hence, we use these spinors as a basis to span the angular part,
\begin{equation}\label{angulardecomp}
\tilde{\psi}=\sum_{\ell m s}\tilde{\psi}_{(+)\ell m}^{(s)}(r,t)\otimes\chi_{(+)\ell m}^{(s)}(\Omega)+\tilde{\psi}_{(-)\ell m}^{(s)}(r,t)\otimes\chi_{(-)\ell m}^{(s)}(\Omega).
\end{equation}
Substituting this ansatz in (\ref{action2}), and integrating out the angular coordinates, we are left with 
\begin{equation}
\begin{split}
S=i\sum_{\ell m s}\int dt dr \sqrt{\tilde{g}_{d=2}}&\left\{\overline{\tilde{\psi}}_{+ \ell m}^{(s)}\left(\tilde{\slashed{\nabla}}_{d=2}-\left(\ell+\frac{d-2}{2}\right)\sigma^1\right)\tilde{\psi}_{+ \ell m}^{(s)}\right.\\
&+\left. \overline{\tilde{\psi}}_{- \ell m}^{(s)}\left(\tilde{\slashed{\nabla}}_{d=2}+\left(\ell+\frac{d-2}{2}\right)\sigma^1\right)\tilde{\psi}_{- \ell m}^{(s)}\right\}\end{split}
\end{equation}
This is a sum of independent free Dirac fermions in two dimensions, with an extra term that breaks chirality. However, in order for the spinor to satisfy canonical anti-commutation relations, we make the weyl transformation
\begin{equation}
\begin{split}
d s_{d=2}^2 &= r^2 d \tilde{s}_{d=2}^2=d t^2 - d r^2\\
{\psi}_{(\pm) \ell m}^{(s)}&=r^{-1/2}\tilde{\psi}_{(\pm) \ell m}^{(s)}
\end{split}
\end{equation}
leading to
\begin{equation}\label{reducedfermion1}
S=\sum_{\ell m s}S_{\ell m s}
\end{equation}
\begin{equation}\label{reducedfermion2}
\begin{split}
S_{\ell m s}=i\int dt dr &\left\{\overline{\psi}_{+ \ell m}^{(s)}\left(\sigma^3\partial_t+i\sigma^2\partial_r -\frac{\left(\ell+\frac{d-2}{2}\right)}{r}\sigma^1\right)\psi_{+ \ell m}^{(s)}\right.\\
&\left.+ \overline{\psi}_{- \ell m}^{(s)}\left(\sigma^3\partial_t+i\sigma^2\partial_r +\frac{\left(\ell+\frac{d-2}{2}\right)}{r}\sigma^1\right)\psi_{- \ell m}^{(s)}\right\}\end{split}
\end{equation}
Note that for each mode, we have twice the Dirac fermion in $d=2$, with a "mass" term that scales as $1/r$. As anticipated, the rescaled spinors have the familiar scaling dimensions and satisfy canonical anti-commutation relations
\begin{equation}
\left\{\psi_{(\pm) \ell m}^{(s)}(r),\psi_{(\pm) \ell' m'}^{(s')\dagger}(r')\right\}=i\delta(r-r')\delta_{\ell'\ell}\delta_{m'm}\delta_{s's}.
\end{equation}
The Hamiltonian of the $d$ dimensional fermion can be dimensionally reduced in an analogous way. We begin with
\begin{equation}
H=\frac{i}{2}\int dr r^{d-2} d\Omega \left( \psi^{\dagger}\partial_t \psi-c.c.\right)\, .
\end{equation}
Writing the spinor as in (\ref{angulardecomp}), and rescaling it with the appropriate $r^{(d-2)/2}$ factor, we finally get
\begin{equation}\label{Hamiltonian}
\begin{split}
H=\frac{i}{2}\sum_{\ell m s}\int dr &\left\{\psi_{+ \ell m}^{(s)\dagger}\left(-\sigma^1\partial_r+\frac{\left(\ell+\frac{d-2}{2}\right)}{r}i\sigma^2\right)\psi_{+ \ell m}^{(s)}- c.c.\right.\\
&\left.+ \psi_{- \ell m}^{(s)\dagger}\left(-\sigma^1\partial_r-\frac{\left(\ell+\frac{d-2}{2}\right)}{r}i\sigma^2\right)\psi_{- \ell m}^{(s)}- c.c.\right\}\end{split}
\end{equation}
The above results are consistent with the dimensional reduction carried out in \cite{Cho:2007zi, Gibbons:2008gg, Das:1996we} to solve the Dirac equation in more general spherically symmetric backgrounds. Equivalent equations of motion are also found in the context of fermions coupled to central potentials \cite{Dong:2003xy, Kanellopoulos:1972jz} where, instead of the Dirac operator eigenspinors in the transverse direction, the authors use spinor spherical harmonics to represent the angular dependence. Besides, equation (\ref{Hamiltonian}) coincides with the dimensionally reduced Hamiltonian reported in \cite{Benedetti:2023jye} and \cite{Huerta:2011qi}, corresponding to four and three dimensions, respectively.

Now we focus on the variational problem of (\ref{reducedfermion}). Firstly, under the redefinition
\begin{equation}
\hat{\psi}_{- \ell m}^{(s)}=\sigma^1 \psi_{- \ell m}^{(s)}
\end{equation}
the action for the hated $``-"$ mode is exactly the same as that of the $``+"$ mode, consequently leading to identical Hamiltonians. Therefore, we will work with these variables and analyse the equation of motion, which both modes have in common.
 
The equation we are left with is \footnote{$\psi$ stands either for $\psi_{+ \ell m}^{(s)}$ or $\hat{\psi}_{- \ell m}^{(s)}$.}
\begin{equation}
\left(\sigma^3\partial_t +i\sigma^2\partial_r-\frac{\lambda}{r}\sigma^1\right)\psi=0\, ,
\end{equation}
where
\begin{equation}
\lambda\equiv \ell+\frac{d-2}{2}\, .
\end{equation}
Conveniently, in this representation the square of the Dirac operator is diagonal. The solution is
\begin{equation}
\psi=A e^{-i k t}\sqrt{r}\begin{pmatrix}
J_{\lambda+1/2}(k r)\\
-i J_{\lambda-1/2}(k r)
\end{pmatrix}+B e^{-i k t}\sqrt{r}\begin{pmatrix}
Y_{\lambda+1/2}(k r)\\
-i Y_{\lambda-1/2}(k r)
\end{pmatrix}\, ,
\end{equation}
which behaves like 
\begin{equation}\label{saddle}
\psi(r\sim 0)=A e^{-i k t}\begin{pmatrix}
r^{\lambda+1}\\
r^{\lambda}
\end{pmatrix}+B e^{-i k t}\begin{pmatrix}
r^{-\lambda}\\
r^{-\lambda+1}
\end{pmatrix}\, .
\end{equation}
Hence, requiring that the $d$ dimensional solution (\ref{angulardecomp}) is a regular, univaluate spinor at the origin, implies $B=0$. This is consistent with imposing Dirichlet boundary conditions for the effective theory in the half line.

\section{Modular Hamiltonian as a charge}\label{sec: modHam}
The sphere modular Hamiltonian for the free fermion in $d$ dimensions is
\begin{equation}
\begin{split}
K&=2\pi\int_{\vert x\vert <R} d^{d-1}x \left(\frac{R^2-r^2}{2R}\right)T_{00}\\
&=i\pi\int_{\vert x\vert <R}d^{d-1}x\left(\frac{R^2-r^2}{2 R}\right)\left(\psi^{\dagger}\partial_t\psi-\partial_t \psi^{\dagger}\psi\right)\, .
\end{split}
\end{equation}
Following the same steps as the ones explained in the previous section for the dimensional reduction of the Hamiltonian, we get
\begin{equation}
K=\sum_{\ell m s}K_{\ell m s}\, ,
\end{equation}
where
\begin{equation}\label{Knls}
\begin{split}
K_{\ell m s}&=2\pi\int_0 ^R dr\left(\frac{R^2-r^2}{2R}\right) T_{0 0}^{\ell m s}\\
&=i\pi\int_0 ^R  dr\left(\frac{R^2-r^2}{2R}\right) \left(\psi_{+\ell m}^{(s)\dagger}\partial_t\psi_{+ \ell m}^{(s)}+\psi_{- \ell m}^{(s)\dagger}\partial_t\psi_{- \ell m}^{(s)}-c.c.\right)\, ,
\end{split}
\end{equation}
Given that the reduced model (\ref{reducedfermion1}) is a sum of independent theories defined on the half line, and that accordingly the density matrix must decompose as a direct product over each sector, it is clear that (\ref{Knls}) is nothing but the modular Hamiltonian for (\ref{reducedfermion2}), restricted to the segment $(0,R)$. The fact that it turns out to be a local expression, proportional to the energy density, is quite remarkable, because the reduced theory is not conformal. However, some portion of the original conformal group survives the dimensional reduction. We will show that the modular transformation belongs to this inherited symmetry group, so that (\ref{Knls}) can be interpreted as the Noether charge acting on the segment \footnote{Here we are interested in the generator of the modular flow on a half diamond, which is the causal development of the segment attached to the origin. The proper conserved Noether charge would be the generator acting on the whole spacetime.}.

On the one hand, recall that under any conformal transformation, satisfying
\begin{equation}
\frac{\partial x'^{\mu}}{\partial x^{\nu}}=\Omega(x) R^{\mu}_{\nu}(x), \quad  R^{\mu}_{\nu}(x) \in SO(1,d-1)
\end{equation}
the fermion transforms \cite{Simmons-Duffin:2016gjk}
\begin{equation}
\psi^{'a}(x')=\Omega(x) ^{-\Delta} D (R(x))_b^a \psi^b (x).
\end{equation}
On the other hand, the infinitesimal modular flow reads, in spherical coordinates
\begin{equation}\label{symmetry}
\begin{split}
& t\longrightarrow t'=t + \epsilon\frac{\pi}{R}(R^2-t^2-r^2)\\
& r\longrightarrow r'=r -2\epsilon\frac{\pi}{R} t r\\
\end{split}
\end{equation}
It is straightforward to check that this is a conformal transformation\footnote{In fact, it is a composition of a time translation and a special conformal transformation, which together with dilatations make up an $SL(2,\mathbb{R})$ symmetry group. See \cite{Huerta:2022tpq}.}, with conformal factor
\begin{equation}
\Omega(t,r)\sim 1-\frac{2\pi\epsilon}{R} t\, ,
\end{equation}
and local boost
\begin{equation}
R^{\mu}_{\nu}(t,r)\sim\delta^{\mu}_{\nu}+\omega ^{\mu}_{\nu}(t,r)\, , \quad \omega_{1 2}(t,r)=\frac{2\pi\epsilon}{R}r\, .
\end{equation}
Therefore\footnote{Once again, here $\psi$ represents either $\psi_{+ \ell m}^{(s)}$ or $\hat{\psi}_{- \ell m}^{(s)}$.}, 
\begin{equation}
\psi'(t',r')=(1+\frac{\epsilon \pi t}{R}+\frac{\epsilon \pi r}{ R}~ \sigma^1)\psi(t,r)+\mathcal{O}(\epsilon^2).
\end{equation}
In order to show that this leaves (\ref{reducedfermion2}) invariant, it is enough to analyse the mass term, since the kinetic term is trivially conformal. Explicitly, we have
\begin{equation}\label{conf_mass}
\int dt dr \overline{\psi} \frac{\lambda}{r}\sigma^1 \psi \rightarrow \int dt dr \Omega^{2}(t,r)\overline{\psi} \frac{\lambda}{r'}\sigma^1 \psi \Omega^{-1}(t,r)\, ,
\end{equation}
but since
\begin{equation}
\frac{1}{r'}\sim\frac{1}{r}\left(1+\frac{2\pi\epsilon}{R}t\right)\sim \frac{\Omega^{-1}(t,r)}{r}
\end{equation}
then both sides of (\ref{conf_mass}) are equal.

The (classically conserved) current associated to the modular transformation is
\begin{equation}
j^{\mu}=\frac{\partial\mathcal{L}}{\partial(\partial_{\mu}\psi)}\Delta\psi +\Delta\overline{\psi}\frac{\partial\mathcal{L}}{\partial(\partial_{\mu}\overline{\psi})}-\mathcal{L}\zeta^{\mu}, 
\end{equation}
where 
\begin{equation}
\Delta\psi=\psi'(t,r)-\psi(t,r)\sim \left(\frac{\epsilon \pi t}{R}+\frac{\epsilon \pi r}{R} \sigma^1 -\zeta^{\nu}\partial_{\nu}\right)\psi(t,r)+\text{subleading}.
\end{equation}
Since the Lagrangian vanishes on shell, we get
\begin{equation}
j^{\mu}=\frac{i}{2}\zeta^{\nu}\left[ \overline{\psi}\gamma^{\mu}\partial_{\nu}\psi-c.c.\right]\, ,
\end{equation}
or, in components
\begin{equation}
j^t=\frac{i\pi}{2 R} \left[(R^2-t^2-r^2)\psi^{\dagger}\partial_t\psi-2tr \psi^{\dagger}\partial_r\psi -c.c.\right]
\end{equation} 
\begin{equation}
j^r=\frac{i\pi}{2 R} \left[(R^2-t^2-r^2)\psi^{\dagger}\sigma^1\partial_t\psi-2tr \psi^{\dagger}\sigma^1\partial_r\psi -c.c.\right].
\end{equation}
This means that the generator of (\ref{symmetry}) on the Hilbert space is
\begin{equation}\label{Kfermion}
\begin{split}
K&=\int_0^R dr j^t(t=0,r)\\
&=i\pi\int_0 ^R  dr\left(\frac{R^2-r^2}{2R}\right) \left(\psi^{\dagger}\partial_t\psi-c.c.\right),
\end{split}
\end{equation}
in perfect agreement with (\ref{Knls})\footnote{This also reduces to the modular Hamiltonian of a free fermion in the half-line reported in \cite{Mintchev:2020jhc} and \cite{Mintchev:2020uom} when $d=2$ and $\ell=0$, as expected.}.
\section{Spectrum of the Modular Hamiltonian}\label{sec: spectrum}
By use of the equations of motion, the modular Hamiltonian (\ref{Kfermion}) can be recast in the form
\begin{equation}
K=2\pi\int_0^R dr \int_0^R dr' \psi^{\dagger}(r') \mathcal{K}(r',r) \psi(r),
\end{equation}
where
\begin{equation}
\mathcal{K}(r',r)= \frac{i}{2}\delta(r'-r)\left[-\left(2\beta(r)\partial_r+\partial_r\beta(r)\right)\sigma^1+\frac{2\lambda }{r}\beta(r)i\sigma^2\right].
\end{equation}
In this section we compute the spectrum of the above kernel, that is, the eigenfunctions satisfying
\begin{equation}\label{eigenKernel}
\int_0^R dr'\mathcal{K}(r,r')\psi_s (r')=s ~\psi_s (r)\, ,
\end{equation}
for $s \in \mathbb{R}$, since $\mathcal{K}$ is an hermitian operator. This leads to the diagonal representation of the modular Hamiltonian
\begin{equation}
K= 2\pi\int_0^R dr\int_{-\infty}^{\infty} ds ~\psi^{\dagger}_s(r)~ s ~\psi_s(r)\, .
\end{equation}
The eigenfunctions are spinors
\begin{equation}
\psi_s (r)=N\begin{pmatrix}
\psi_{s u}(r) \\
\psi_{s d}(r)
\end{pmatrix}\, ,
\end{equation}
with components
\begin{equation}\label{psiu1}
\psi_{s u}(r)= \frac{- R^{1/2}s \sinh{(2\pi s)}}{\pi \Gamma[3/2+\lambda]}\left(\frac{r}{R}\right)^{\lambda+1}\beta(r)^{-\frac{1}{2}-i s}\left(\frac{R}{2}\right)^{i s} {}_2F_1\left[1-i s,\frac{1}{2}-i s+\lambda,\frac{3}{2}+\lambda,\frac{r^2}{R^2}\right]\, ,
\end{equation}
\begin{equation}\label{psid1}
\psi_{s d}(r)=\frac{i R^{1/2} \sinh{(2\pi s)}}{\pi \Gamma[1/2+\lambda]}\left(\frac{r}{R}\right)^\lambda\beta(r)^{-\frac{1}{2}-i s}\left(\frac{R}{2}\right)^{i s}{}_2F_1\left[-i s,\frac{1}{2}-i s+\lambda,\frac{1}{2}+\lambda,\frac{r^2}{R^2}\right]\, ,
\end{equation}
and $N$ a normalization factor.
These are regular eigenfunctions which behave as
\begin{equation}
\psi_s(r)\sim \begin{pmatrix}
r^{\lambda+1}\\
r^{\lambda}
\end{pmatrix}
\end{equation}
near the origin, in agreement with the saddle point configuration (\ref{saddle}). An independent set of eigenfunctions was dismissed due to being divergent at $r= 0$. 

Meanwhile, near the boundary
\begin{equation}\label{psiumas}
\psi_{s u} (r)\sim \left(\frac{R-r}{R}\right)^{-\frac{1}{2}-i s}\alpha(s)+ c.c.
\end{equation}
\begin{equation}\label{psidmas}
\psi_{s d}(r)\sim \left(\frac{R-r}{R}\right)^{-\frac{1}{2}-i s}\alpha(s)- c.c.\, ,
\end{equation}
where we define 
\begin{equation}
\alpha(s)=\frac{ 2^{-i s}}{\Gamma\left[1-2 i s\right]\Gamma\left[i s\right]\Gamma\left[\frac{1}{2}+i s+\lambda\right]}\, .
\end{equation}
In this limit the properties
\begin{equation}\label{parity1}
\psi_{s u}^{*}(r)=\psi_{s u}(r)=\psi_{-s u}(r)
\end{equation}
\begin{equation}\label{parity2}
\psi_{s d}^{*}(r)=-\psi_{s d}(r)=\psi_{-s d}(r)\, ,
\end{equation}
become manifest. The fact that one component must be real and the other imaginary can be straightforwardly deduced from the differential equations. 

Now we compute the normalization factor $N$, by requiring that
\begin{equation}\label{orthogonality}
\int_0^R dr \psi_s ^{\dagger}(r)\psi_{s'}(r)= \delta(s-s')\, .
\end{equation}
As the only piece capable of contributing with a Dirac delta function is the one near the boundary $r\sim R$, we substitute (\ref{psiumas}) and (\ref{psidmas}), leading to
\begin{equation}
N=\frac{1}{\sqrt{4\pi R} \vert\alpha(s)\vert}\, ,
\end{equation}
save an overall phase that we set to one for simplicity.

\section{The entropy}\label{sec: entropy}
The advantage of having calculated the spectrum of the modular Hamiltonian (\ref{Kfermion}) is that it allows us to work out an analytic expression for the entanglement entropy of a segment, corresponding to the vacuum of each one dimensional theory (\ref{reducedfermion2}). We devote the first part of this section to derive this expression. Later, we sum over the independent modes to recover the entanglement entropy of the $d$ dimensional Dirac fermion in the sphere, paying special attention to the universal terms.
\subsection{Entropy of the effective theories in the segment}
\begin{figure}[t]
\begin{center}  
\includegraphics[width=0.50\textwidth]{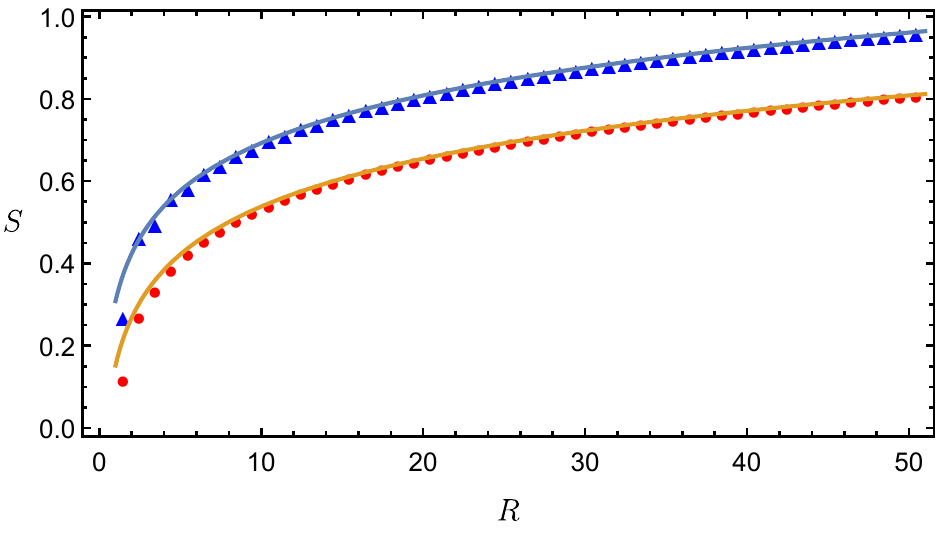}
\captionsetup{width=0.9\textwidth}
\caption{$S_{+}(\ell,d=3)$ as a function of the disk radius. The blue triangles correspond to $\ell=1$ and the red circles to $\ell=0$. We used a lattice of total size $m=3000$. The curves represent the best fit with a function $S(R)=c_{0}+c_{\log}\log{R}$. We find $c_{\log}(\ell=0)\sim 0.167$ and $c_{\log}(\ell=1)\sim 0.168$, in agreement with (\ref{Smas}). }
\label{SvsR}
\end{center}  
\end{figure}

Since the one dimensional effective theory (\ref{reducedfermion2}) is a sum, the entropy of a segment $(0, R)$ naturally reads
\begin{equation}\label{TotalEntropy}
S(\ell,d)=S_{+}(\ell,d)+S_{-}(\ell,d)\, ,
\end{equation}
where the two terms on the right hand side contribute equally, because the modular Hamiltonians corresponding to the $``+"$ and the $``-"$ modes coincide. Using the eigenfunctions (\ref{psiu1}) and (\ref{psid1}), we get
\begin{equation}
S_{\pm}(\ell,d)=\int_0^{R-\epsilon} dr \int_{-\infty}^{\infty} ds ~\psi_s^{\dagger}(r) f(s)\psi_{s}(r)\, .
\end{equation}
The function $f(s)$ is the entropy density \cite{Casini:2009sr}
\begin{equation}
f(s)=\log{\left(1+e^{-2\pi s}\right)}+2\pi s\frac{e^{-2\pi s}}{1+e^{-2\pi s}}\, .
\end{equation}
Note that since we expect the entropy to diverge due to the short range correlations between degrees of freedom at both sides of the boundary, we regularized the above integral by introducing a small cutoff $\epsilon$. After invoking the symmetry properties of the spinors under parity transformation of the eigenvalue $s$ (\ref{parity1}), (\ref{parity2}), which make the integral parity even, we also make one further regularization,
\begin{equation}
S_{\pm}(\ell,d)=2\lim\limits_{\delta s\rightarrow 0}\int_0^{R-\epsilon} dr \int_0^{\infty} ds ~\psi_s^{\dagger}(r) f(s)\psi_{s+\delta s}(r)\, , 
\end{equation}
which consists of summing over slightly off diagonal elements by shifting $s\rightarrow s+\delta s$ \cite{Arias:2018tmw}. This is very convenient, because for fixed $\delta s\neq 0$, the orthogonality relation (\ref{orthogonality}) allows us to change the domain of integration, according to
\begin{equation}
\begin{split}
\int_0^{R-\epsilon} dr \psi_s^{\dagger}(r)\psi_{s+\delta s}(r)&=\int_0^{R} dr \psi_s^{\dagger}(r)\psi_{s+\delta s}(r)-\int_{R-\epsilon}^{R} dr \psi_s^{\dagger}(r)\psi_{s+\delta s}(r)\\
&=-\int_{R-\epsilon}^{R} dr \psi_s^{\dagger}(r)\psi_{s+\delta s}(r)\, .
\end{split}
\end{equation}
Hence
\begin{equation}
S_{\pm}(\ell,d)=-2\lim\limits_{\delta s\rightarrow 0}\int_{R-\epsilon}^{R} dr \int_0^{\infty} ds ~\psi_s^{\dagger}(r) f(s)\psi_{s+\delta s}(r)\, . 
\end{equation}
Crucially, we can now substitute the Taylor expansion near the boundary (\ref{psiumas}), (\ref{psidmas}), that is much easier to integrate than the original spinor (\ref{psiu1}), (\ref{psid1}). A straightforward computation yields,
\begin{equation}
\int_{R-\epsilon}^{R} dr \psi_s^{\dagger}(r)\psi_{s+\delta s}(r)=\frac{1}{\pi}\log{\frac{\epsilon}{R}}+\frac{i}{2\pi}\left(\frac{\alpha'(s)}{\alpha(s)}-\frac{\alpha^{* '}(s)}{\alpha^{*}(s)}\right)+\mathcal{O}(\delta s)\, ,
\end{equation}
so, finally
\begin{equation}\label{Smas}
S_{\pm}(\ell,d)=\frac{1}{6}\log{\frac{R}{\epsilon}}-\frac{2}{\pi}\int_0^{\infty} ds f'(s) \text{Arg}\left(\alpha(s)\right)\, .
\end{equation}
The logarithmic coefficient is the expected one for a free Dirac fermion in the half line \cite{Cardy:2016fqc}, that is, $c/6$, where $c$ the Virasoro central charge and each chirality contributes with $c=1/2$. The result of numerical computations in the radial lattice, which we will explain at the end of this section, is shown in figure (\ref{SvsR}).

\begin{figure}[t]
\begin{center}  
\includegraphics[width=0.50\textwidth]{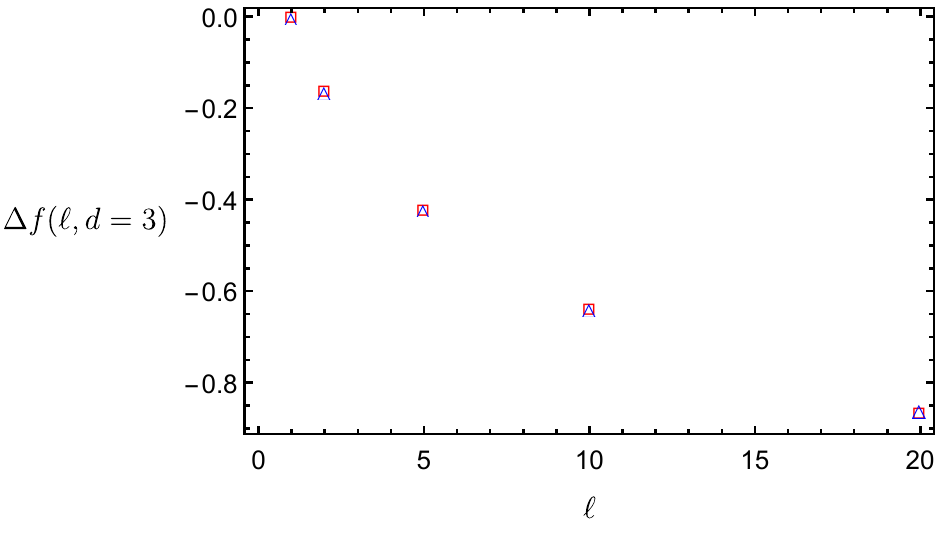}
\captionsetup{width=0.9\textwidth}
\caption{$\Delta f(\ell,d=3)\equiv f(\ell,3)-f(\ell=1,3)$. The blue triangles represent numerical computation in the radial lattice, whereas the red squares correspond to direct numerical evaluation of the integral (\ref{fld}). The lattice size used is $m=2000$, with fixed disk radius $R=200$.}
\label{num_f_l_d3_paper}
\end{center}  
\end{figure}

Summing the two independent pieces in (\ref{TotalEntropy}) we arrive at\footnote{Note that at $d=2$ there is no decomposition into $\pm$ modes, so we consistently recover $S(d=2)=\frac{1}{6}\log{\frac{R}{\epsilon}}$.}
\begin{equation}\label{Sld_completa}
S(\ell,d)=\frac{1}{3}\log{\frac{R}{\epsilon}} - 2 i \pi \int_0^{\infty} ds \frac{s}{\cosh^2{\pi s}}\log{\left[\frac{2^{-2 i s}\Gamma\left(1+2 i s\right)\Gamma\left(-i s\right)\Gamma\left(\ell+\frac{d-1}{2}-i s\right)}{\Gamma\left(1-2 i s\right)\Gamma\left(i s\right)\Gamma\left(\ell+\frac{d-1}{2}+i s\right)}\right]}\, ,
\end{equation}
or 
\begin{equation}
S(\ell,d)=\frac{1}{3}\log{\frac{R}{\epsilon}}+f(\ell,d)+q\, ,
\end{equation}
where we conveniently separated the constant term into a piece that depends both on the mode $\ell$ and the spacetime dimension of the parent theory $d$,
\begin{equation}\label{fld}
f(\ell,d)\equiv - 2 i \pi \int_0^{\infty} ds \frac{s}{\cosh^2{\pi s}}\log{\left[\frac{\Gamma\left(\ell+\frac{d-1}{2}-i s\right)}{\Gamma\left(\ell+\frac{d-1}{2}+i s\right)}\right]}
\end{equation}
and another piece which does not depend on any of those parameters
\begin{equation}
q\equiv -2 i \pi \int_0^{\infty} ds \frac{s}{\cosh^2{\pi s}}\log{\left[\frac{2^{-2 i s}\Gamma\left(1+2 i s\right)\Gamma\left(-i s\right)}{\Gamma\left(1-2 i s\right)\Gamma\left(i s\right)}\right]}\, .
\end{equation}
In figure (\ref{num_f_l_d3_paper}) we compare the value of $f(\ell,d=3)$ numerically calculated by different means: either regularizing the effective one dimensional Hamiltonian in a radial lattice (for a brief review on this, see next subsection), or directly evaluating (\ref{fld}). Since the constant term depends on the regularization scheme, we substract the one corresponding to $\ell=1$ and compare $\Delta f(\ell,3)\equiv f(\ell,3)-f(\ell=1,3)$.

\begin{figure}[t]
\begin{center}  
\includegraphics[width=0.50\textwidth]{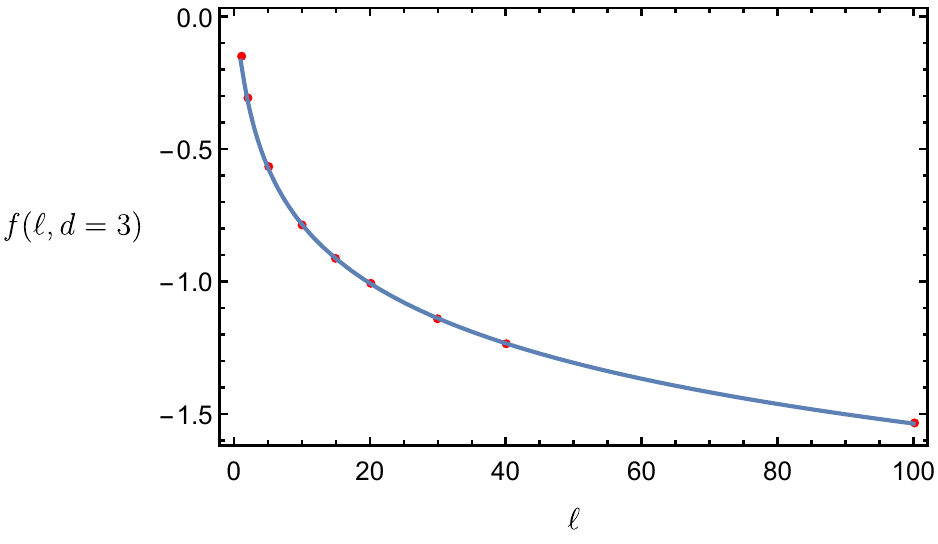}
\captionsetup{width=0.9\textwidth}
\caption{Constant contribution of the entropy $f(\ell,d=3)$, as a function of the angular mode $\ell$. Red dots: numerical computation of (\ref{fld}), for the set $\ell=\{1,2,5,10,15,20,30,40,100\}$. Blue curve: $f(\ell)=-\frac{1}{3}\log{\ell}-\frac{1}{6\ell}$, the leading terms in the large $\ell$ approximation (\ref{flarge}).}
\label{fld3}
\end{center}  
\end{figure}

Although the integral in (\ref{fld}) cannot be solved analytically as a function of $\ell$ and $d$, at sufficiently large $\ell>>1$ we can approximate
\begin{equation}\label{flarge}
f(\ell,d)\sim -\frac{1}{3}\log{\ell}+\frac{a_1}{\ell}+\frac{a_2}{\ell^2}+\frac{a_3}{\ell^3}+\frac{a_4}{\ell^4}+\mathcal{O}\left(\frac{1}{\ell^5}\right)\, ,
\end{equation}
where
\begin{eqnarray}\label{a1}
a_1&=&\frac{1}{3}-\frac{d}{6}\\ \label{a2}
a_2&=&\frac{2}{15}-\frac{d}{6}+\frac{d^2}{24}\\ \label{a3}
a_3&=&\frac{2}{45}-\frac{2 d}{15}+\frac{d^2}{12}-\frac{d^3}{72}\\ \label{a4}
a_4&=&-\frac{1}{210}-\frac{d}{15}+\frac{d^2}{10}-\frac{d^3}{24}+\frac{d^4}{192} 
\end{eqnarray}
The first two terms in the above expansion already approximate $f(\ell, d)$ at $\ell\sim \mathcal{O}(1)$ very accurately, as shown in figure (\ref{fld3}).
\subsection{Universal terms in the entropy of the free fermion in the sphere}
As explained in section \ref{sec: modHam}, the modular Hamiltonian of the free Dirac fermion in the sphere $r< R$ decomposes into independent modular Hamiltonians, labelled by the set $\{n, \ell, s\}$, each representing the modular Hamiltonian of an effective one-dimensional theory in the segment $(0,R)$. Therefore, we expect that after summing the entropies obtained in the previous section over $\{n, \ell, s\}$ we must be able to reproduce the general structure
\begin{equation}\label{Sgen}
S=\begin{cases}
      \#\left(\frac{R}{\epsilon}\right)^{d-2}+ ... +c_{\text{log}} \log{\frac{R}{\epsilon}}, & d \quad\text{even} \\
       \#\left(\frac{R}{\epsilon}\right)^{d-2}+ ... + F. & d \quad \text{odd}
    \end{cases}
\end{equation}
In even dimensions the universal quantity is the logarithmic coefficient, which is the trace anomaly coefficient associated to the Euler density, whereas in odd dimensions the universal piece is the the constant term, which is the free energy of the euclidean theory in a sphere.

From (\ref{flarge}) we learn that $S(\ell,d)$ grows very rapidly as a function of $\ell$, so we need to regularize the sum. We do so by introducing a damping exponential,
\begin{equation}\label{Ssum}
S=\sum_{\ell=0}^{\infty}\lambda(\ell,d)S(\ell,d)e^{-\left(\ell+\frac{1}{2}\right)\epsilon/R}\, ,
\end{equation}
with $\lambda(\ell,d)$ the density of states
\begin{equation}\label{lambda}
\lambda(\ell,d)=\frac{2^{\left[\frac{d-2}{2}\right]}(d+\ell-3)!}{(d-3)!\ell!} \, .
\end{equation}
Note that the exponent in (\ref{Ssum}) is shifted $\ell\rightarrow \ell+1/2$ when compared to the one put forward in the scalar case \cite{Huerta:2022tpq}. We will delve into this later.

Since we are unable to analytically compute the integral (\ref{Sld_completa}) that defines $S(\ell,d)$, in (\ref{Ssum}) we substitute it by its large $\ell$ expansion,
\begin{equation}
S=\sum_{\ell=1}^{\infty}\lambda(\ell,d)\left( \frac{1}{3}\log{\frac{R}{\epsilon}}-\frac{1}{3}\log{\ell} +\sum_{j=1}^{j_{max}}\frac{a_j}{\ell^j}+ q\right)e^{-\left(\ell+\frac{1}{2}\right)\epsilon/R}+\lambda(0,d)S(0,d)+\text{correction}\, .
\end{equation}
The correction above accounts for the error made in the approximation,
\begin{equation}
\text{correction}=\lim_{\ell_{max}\rightarrow \infty}\sum_{\ell=1}^{\ell_\text{max}}\left(f(\ell,d)+\frac{1}{3}\log{\ell}-\sum_{j=1}^{j_{max}}\frac{a_j}{\ell^j}\right)\, .
\end{equation}
On the other hand, the power series in $1/\ell$ is truncated at $j=j_{max}$. To make sense of the sum, we must include at least as many terms as those which require regularization, that is, $j_{max}\geq d-2$. Subleading terms give finite contributions, and might already be taken into account by the correction. In fact, the divergent pieces come from terms of the general form
\begin{equation}\label{divS1}
\sum_{\ell=1}^{\infty} \ell^p \left( \log{\frac{R}{\epsilon}}-\log{\ell}\right)e^{-\ell \epsilon/R}= -\Gamma'(p+1)\left( \frac{R}{\epsilon}\right)^{p+1}+\zeta(-p)\log{\frac{R}{\epsilon}}+\zeta'(-p),
\end{equation}
\begin{equation}\label{divS2}
\sum_{\ell=1}^{\infty} \ell^p e^{-\ell \epsilon/R}= p!\left( \frac{R}{\epsilon}\right)^{p+1}+\zeta(-p),
\end{equation}
with $\{p\mid p\in\mathbb{N}_0\wedge p\leq d-3\}$, and
\begin{equation}\label{divS3}
\sum_{\ell=1}^{\infty} \frac{1}{\ell}e^{-\ell \epsilon/R}= \log{\frac{R}{\epsilon}}\, .
\end{equation}
so the fact that (\ref{lambda}) grows as $\ell^{d-3}$ implies that $j_{max}\geq d-2$.

In even dimensions, the contributions to the logarithmic coefficient come from (\ref{divS1}) and (\ref{divS3}). Just to list some relevant specific cases, we get	
\begin{equation}\label{c4}
c_{\text{log}}(d=4)=\frac{2}{3}\left(\zeta(-1)+\zeta(0)+1\right)+2\left(a_1+a_2\right)=-\frac{11}{90}\, ,
\end{equation}
\begin{equation} \label{c6}
c_{\text{log}}(d=6)=\frac{2}{3}\left(\frac{1}{3}\zeta(-3)+\frac{11}{3}\zeta(-1)+2\zeta(0)+2\right)+2\left(2 a_1+\frac{11}{3}a_2+2 a_3+\frac{1}{3}a_4\right)=\frac{191}{3780}\, ,
\end{equation}
\begin{equation}\label{c8}
c_{\text{log}}(d=8)=\ldots=-\frac{2497}{113400}\, .
\end{equation}
These agree with the values reported in the literature \cite{Cappelli:2000fe}. Besides, we can check that the logarithmic coefficient vanishes in odd dimensions, as expected.

\begin{figure}[t]
\begin{center}  
\includegraphics[width=0.50\textwidth]{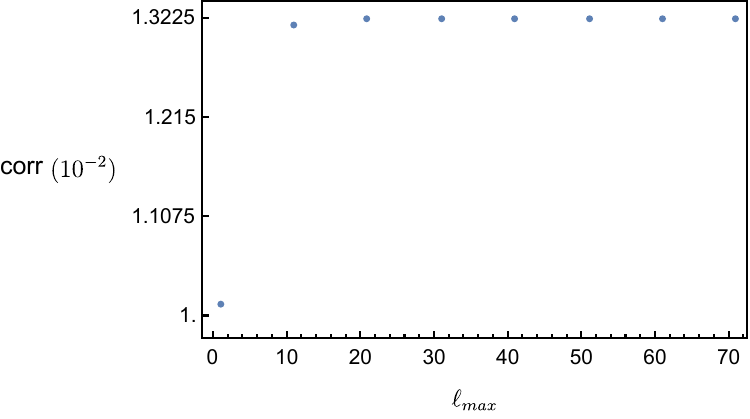}
\captionsetup{width=0.9\textwidth}
\caption{Correction computed at $d=3$, as a function of $\ell_{max}$. }
\label{correction_plot}
\end{center}  
\end{figure}

Regarding the constant $F$ term in three dimensions, we take $j_{max}=2$, yielding
\begin{equation}
F_{d=3}=\frac{1}{3}\zeta'(0) +a_2 \zeta(2)+ f(0,d=3)+\frac{\Gamma'(1)}{6}+ \text{correction}\, ,
\end{equation}
where in this case
\begin{equation}
\text{correction}=\lim_{\ell_{max}\rightarrow \infty}\sum_{\ell=1}^{\ell_{max}}\left(f(\ell,d=3)+\frac{1}{3}\log{\ell}-\frac{a_1}{\ell}-\frac{a_2}{\ell^2}\right)
\end{equation}
In figure (\ref{correction_plot}) we plot the correction as a function of $\ell_{max}$, and show that it converges rather fast to the value $\sim 0.013225$. Substituting (\ref{a2}) and
\begin{equation}
f(0, 3)\sim 0.15662\, , 
\end{equation}
we finally arrive at
\begin{equation}\label{ourF}
F_{d=3}\sim -0.21896\, .
\end{equation}
The exact value is $\frac{1}{2^3}\left(2\log{2}+3\frac{\zeta(3)}{\pi^2}\right)=-0.21895948...$ \cite{Klebanov:2011gs}.

It is worth stressing that had we chosen a different regularization for the angular modes in (\ref{Ssum}), for example the damping exponential $\exp{(-\ell \epsilon/R)}$ employed for the scalar \cite{Huerta:2022tpq}, then we would have obtained an incorrect $F$ term. Indeed, let us call $F_{s}$ the constant term that results from the regularization scheme defined by $\exp{(-(\ell+s)\epsilon/R)}$, with $s$ an arbitrary constant \footnote{Although we label it with the same letter, this constant should not be confused with the eigenvalue of the modular Hamiltonian kernel (\ref{eigenKernel}).}. It is clear that the following relation holds 
\begin{equation}
F_{s}=F_{s=0}- s \times c_{\text{area}}\, ,
\end{equation}
where $c_{\text{area}}$ is the coefficient of the area term, that proportional to $R/\epsilon$. In particular, the fact that $s=1/2$ leads to the proper universal quantity for the Dirac Fermion, just as $s=0$ is the appropriate value for the scalar, seems to suggest that the correct exponent is the total angular momentum $j=\ell+s$, $s$ representing the spin of the particle.

The same subtlety arises in the lattice calculation of the disk entanglement entropy in $d=3$ dimensions, which we very briefly describe here to illustrate the issue. We follow \cite{Huerta:2011qi}.

\begin{figure}[t]
\begin{center}  
\includegraphics[width=0.50\textwidth]{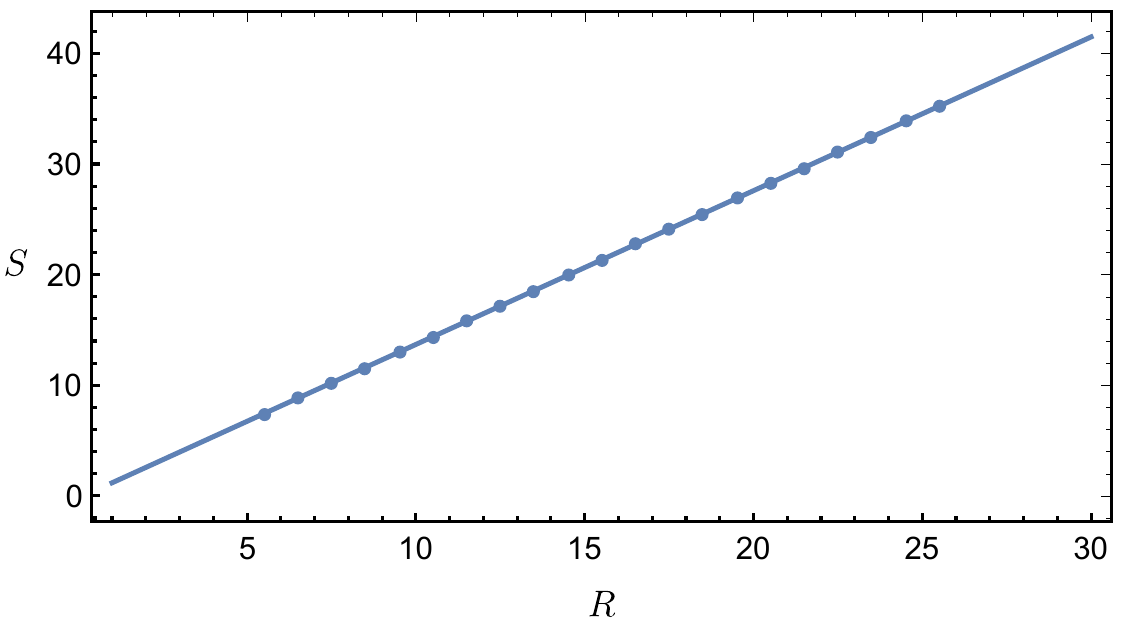}
\captionsetup{width=0.9\textwidth}
\caption{Entanglement entropy as a function of the disk radius. The total lattice size considered is $m=200$. The curve represents a linear fit, with intercept $c_0=-0.21029$. }
\label{Fterm}
\end{center}  
\end{figure}

The lattice regularization of the Hamiltonian for each mode in (\ref{Hamiltonian}) reads
\begin{equation}
H=\sum_{k,j=1}^{m}\sum_{\alpha,\beta=1}^2\psi^{\alpha *}_{k} M_{k,j}^{\alpha,\beta} \psi_{j}^{\beta}\, ,
\end{equation}
where $m$ is the lattice size and
\begin{eqnarray}
M_{k,j}^{1,1}&=&M_{k,j}^{2,2}=0\, ,\\
M_{k,j}^{1,2}&=&i\left[\frac{1}{2}\left(\delta_{k,j+1}-\delta_{k,j-1}\right)+\frac{\lambda}{k}\delta_{k,j}\right]\, ,\\
M_{k,j}^{1,2}&=&i\left[\frac{1}{2}\left(\delta_{k,j+1}-\delta_{k,j-1}\right)-\frac{\lambda}{k}\delta_{k,j}\right]\, ,
\end{eqnarray}
with
\begin{equation}
\lambda	=\ell+\frac{1}{2}\, .
\end{equation}
As explained in \cite{Huerta:2011qi}, from the $2m \times 2m$ matrix with components
\begin{equation}
\tilde{M}^{2 k+\alpha-2, 2 j+\beta-2}=M_{k,j}^{\alpha,\beta}\, ,
\end{equation}
we can compute the correlators $\tilde{C}^{2k+\alpha-2,2j+\beta-2}\equiv \langle \psi_k^{\alpha}\psi^{\beta\dagger}_j\rangle$, according to
\begin{equation}
\tilde{C}=\theta(-\tilde{M})\, ,
\end{equation}
since at zero temperature the fermions follow the Fermi-Dirac distribution. In turn, the entanglement entropy of an interval $(0,R)$ is given by the above correlation matrix, but restricted to the first $2 n$ elements, with $R=n-1/2$,
\begin{equation}
S_R=-\text{Tr}\left[\left(1-\tilde{C}\vert_{2 n}\right)\log{\left(1-\tilde{C}\vert_{2 n}\right)}+\tilde{C}\vert_{2 n}\log{\tilde{C}\vert_{2 n}}\right]\, .
\end{equation} 
Finally, in order to calculate the entanglement entropy of a disk of radius $R$ we must sum the above entropy over each set $\{\ell, m, s\}$. The way in which we account for the contribution of the largest modes, say $\ell>3000$, is by computing the entropy for $\ell=\{5000, 7000, 9000, 11000, 13000, 15000\}$ and then numerically summing the values of the function
\begin{equation}
S_R(\ell)\sim \frac{d_1}{\ell^2}+\frac{d_2}{\ell^2}\log{\ell}+\frac{d_3}{\ell^4}+\frac{d_4}{\ell^4}\log{\ell}+\frac{d_5}{\ell^6}+\frac{d_6}{\ell^6}\log{\ell}
\end{equation} 
that best approximates the data. As shown in figure (\ref{Fterm}), the entropy as a function of the disk radius is linear $S(R)=c_0+c_{\text{area}}R$, with an intercept
\begin{equation}
c_0=-0.21029\, .
\end{equation}
This is a numerically stable result, which is consistent with (\ref{ourF}) given the numerical precision involved. Note that the choice $R=n-1/2$ is crucial in determining $F$. This corresponds to a physically meaningful geometrical regularization, as discussed in \cite{Casini:2015woa}.

\section{Final remarks}

In this article we extended the catalogue of theories with analytic and local modular Hamiltonians. We showed that a Gaussian, non-conformal Dirac fermion defined in the half line, has a modular Hamiltonian proportional to its energy density, with proportionality factor $\beta(r)$, alike the modular Hamiltonian of CFTs in spheres. The theory is obtained from the dimensional reduction of the free fermion in $d$ dimensions, and has a mass term which scales as $1/r$, labelled both by $d$ and the angular modes $\ell$. Although the standard requirement to derive the sphere modular Hamiltonian from the vacuum state on the Rindler wedge is conformal symmetry, the presence of the mass term does not prevent the modular transformations acting in an interval attached to the origin $(0,R)$ from being symmetries. In fact, we proved that the modular Hamiltonian is the corresponding generator. 

Furthermore, we found the eigenfunctions of the resulting modular Hamiltonian, and used them to analytically compute the entanglement entropy of the segment $(0,R)$. This has a logarithmic term with coefficient $c_{\log}=1/6$, as expected for a Dirac fermion in a half-line, together with a constant term which depends on $\ell$ and $d$. We compared this result with numerical calculations in the radial lattice, finding excellent agreement.

Since the vacuum state in a sphere $R$ for the parent $d$ dimensional fermion decomposes as a direct product $\otimes \rho_{\ell m s}$, each factor representing the state of the reduced theory in the interval $(0,R)$, we summed the above entropies over the angular modes to recover the universal terms, namely the anomaly coefficients in even $d$ dimensions and the constant $F$ term in $d=3$. In order to regularize the sum, we introduced a damping exponential $\exp{\left[-(\ell+1/2)\epsilon/R\right]}$. Comparison with the scalar case suggests that the exponent is the total angular momentum $j=\ell+s$. Meanwhile, in odd dimensions $d>3$ the regularization employed does not lead to the correct behaviour of the entanglement entropy. Even powers of $R/\epsilon$ are present in the expansion, meaning that the regularization must be adjusted with the dimension in order to become a geometric one. It would be worth exploring this issue in the future.

Finally, to complete this novel class of modular Hamiltonians in the semi-infinite line, the vector fields would naturally be the next to consider. Nevertheless, as shown in \cite{Casini:2015dsg}, Maxwell theory in $d=4$ turns out to be equivalent to two free scalar fields, the only difference being that the mode $\ell=0$ is removed from the angular decomposition. In $d\neq 4$ dimensions we expect the reduced modular Hamiltonians to be related to the scalar's as well, but given the lack of conformal symmetry the identification might be less obvious. In this regard, it would be interesting to study modular Hamiltonians for theories with generalised symmetries. One example is precisely Maxwell theory in $2+1$ dimensions, which is dual to a free scalar. However, only a portion of the operator algebra is available, more specifically, the one invariant under the global shift symmetry. So the algebra one needs to consider is that of the derivatives of the scalar. It would be relevant to understand how this information is codified in the modular Hamiltonian, and how it in turn leads to the universal terms of the mutual information which signal the presence of generalised symmetries.

\section*{Acknowledgements}
We thank V. Benedetti, G. Hansen and G. Torroba for discussions while this work was being carried out. This work was supported by CONICET, CNEA and Universidad Nacional de Cuyo, Instituto Balseiro, Argentina.

\bibliography{refs}

\bibliographystyle{utphys}

\end{document}